\documentclass[fleqn,twoside]{article}
\usepackage[headings]{espcrc2}

\readRCS
$Id: espcrc2.tex,v 1.2 2004/02/24 11:22:11 spepping Exp $
\usepackage{graphicx}
\usepackage[figuresright]{rotating}

\def\fig#1{Fig.~\ref{#1}}
\def\eq#1{Eq.~(\ref{#1})}

\def\tev{\mbox{~TeV}}

\def\gev{\mbox{~GeV}}
\def\gevc{\mbox{~GeV/$c$}}

\def\mevc{\mbox{~MeV/$c$}}

\def\la{\left< }
\def\ra{\right> }
\def\mean#1{\ensuremath{\la#1\ra}}

\def\meankv#1{\ensuremath{\la#1^2\ra}}
\def\rms#1{\meankv{#1}}
\def\sqrtrms#1{\ensuremath{\sqrt{\meankv{#1}}}}

\def\eg{{\it e.g.}}

\newcommand{\raa} {\ensuremath{R_{\rm AA}}}

\newcommand{\prob}{\ensuremath{\mathcal{P}}}

\newcommand{\auau} {\ensuremath{Au+Au}}
\newcommand{\dau} {\ensuremath{d+Au}}
\newcommand{\pp} {\ensuremath{p+p}}

\newcommand{\ptaf} {\ensuremath{\mathbf{\ptq{a}}}}
\newcommand{\pttf} {\ensuremath{\mathbf{\ptq{\gamma}}}}

\def\ptq#1{\ensuremath{\hat{p}_{T\rm #1}}}

\def\pt#1{\ensuremath{p_{T\rm #1}}} 
 
\def\vpt#1{\ensuremath{\vec{p}_{T\rm #1}}} 
\def\ptg{\ensuremath{p_{T\gamma}}} 

\def\kt#1{\ensuremath{k_{T\rm #1}}}

\def\ptpair{\ensuremath{\mean{\pt{}}_{pair}}}

\newcommand{\zt} {\ensuremath{z_{\rm t}}}

\newcommand{\za} {\ensuremath{z_{\rm a}}}

\newcommand{\xe} {\ensuremath{x_{\rm E}}}

\newcommand{\s} {\ensuremath{\sqrt{s}}}
\newcommand{\piz} {\ensuremath{\pi^0}}

\def\bgi{\begin{itemize}}
\def\endi{\end{itemize}}
\def\bge{\begin{equation}}
\def\ende{\end{equation}}
\def\bgc{\begin{center}}
\def\endc{\end{center}}

\title{From RHIC to LHC}

\author{J. Rak\address[JYFL]{
Department of Physics,\\ 
P.O.Box 35  Jyv\"askyl\"a\\
FI-40014 University of Jyv\"askyl\"a, Finland\\
}}

\begin{document}

\begin{abstract}
Measurements of the nuclear modification factor of inclusive yields of $\pi^{0}$, $p\bar{p}$,
non-phtonic electrons and prompt photon measured at RHIC are reviewed. Some of the
difficulties arising from interpretation of these measurements such as similarity of the
suppression pattern of light and heavy quarks, quarks and gluons, light quarks and prompt
photon are discussed. The potential of two-particle and promt photon correlation technique to
unravel some of these open questions is presented.  An emphasis is given to the influence 
of partonic transverse momentum on the prompt-photon correlations. The smearing between 
the trigger photon and back-to-back jet at LHC energy is discussed.
\vspace{1pc} \end{abstract}

\maketitle

\section{Introduction}

A vast number of experimental results from the Relativistic Heavy Ion Collider (RHIC) changed,
in many aspects, our understanding of ultra-relativistic nuclear collisions.  High-\pt{}\
particle suppression observed in central \auau\ collisions (\eg\ \cite{Adler:2006hu}) and
its absence in \dau\ collisions (\eg\ \cite{Adler:2006wg}) where no excited nuclear medium
exists provides quite firm indication of the final state interaction induced by a deconfined
QCD medium \cite{PHENIX_white}.  Excessive production of baryons at intermediate \pt{}\
\cite{Adler:2003cb} is often interpreted as another manifestation on the deconfined medium.
The dense partonic medium does not hadronize in the usual way of jet fragmentation, but partons
rather recombine forming the baryon wave-function directly.  This so called
``coalescence'' mechanism \cite{Fries:2003kq} is quite successful in explaining many of
the RHIC observations like ordering of azimuthal anisotropy parameter $v_{2}$
(\cite{Lacey:2006bc} and references therein).

Although the theoretical interpretation of above mentioned phenomena is quite successful,
many open questions still remain to be answered. Some of the most important are 
how exactly partons are loosing their energy when they interact with deconfined QCD
medium, where is this energy going to, how does the QCD medium modify the parton properties,
fragmentation process and many others.

\section{Inclusive yields}

Most successful models of partonic energy loss (GLV \cite{Gyulassy:2000fs}, BDMPS
\cite{Baier:1996sk}, Wang \cite{Wang_NonAbelianQuenching} and others) are exploring the
inelastically induced gluon radiation process.  These models are quite successful in
describing the nuclear modification factor \raa\ defined as ratio of inclusive particle
yield in heavy ion collision to the binary collision scaled yield in \pp\ data.
However,  T. Renk and K. Eskola in \cite{Renk:2006pk} studied the sensitivity of
the \raa\  parameter to the details of parton interaction with QCD medium. They argued
that due to the strong quenching the particles reaching the detector may be radiated from
the jets originating close to the surface of the QCD medium and those produced deep in the
medium are fully absorbed.  In this case the sensitivity of \raa\  to the detailed modeling of the 
parton energy loss is reduced. In this light it is not surprising that
predictions from other models \eg\  \cite{Borghini:2005em} introducing 
different mechanism of gluon radiation are also able to reproduce measured values of \raa.
It is evident that more complex observables like two-particle correlations or
reconstructed jets properties has to be explored.

Another open question arose from the analysis of the non-photonic electron yield
originating from heavy (charm and bottom) quark fragmentation
\cite{Adler:2004ta,Adler:2005xv,Bielcik:2007ii}. Induced radiation from heavy
non-relativistic quarks is expected to be weakened by the dead-cone effect
\cite{Dokshitzer:2001zm}. However, high-\pt{}\ $e^{\pm}$ suppression seems to be as
strong as it is in the case of light, $u$ and $d$ quarks. One of the possible explanations
is that quarks are losing a significant fraction of their energy also via elastic
2$\rightarrow$2 processes \cite{Wicks:2005gt} often referred as collisional energy loss.
Since the collisional energy loss plays obviously more important role for heavier partons
it can compensate for the weaker radiational energy loss. However, this mechanism should
be accompanied by significant broadening of the away-side hadron azimuthal distribution
\cite{Vitev:2003jg,Vitev:2005_LargeAngleCorrel} which is not yet seen in the data \cite{Adams:2006yt}.
An alternative model of heavy quark quenching, which does not require any broadening,
attributes the quenching pattern to the hadronic rather to partonic interaction
\cite{Adil:2006ra}.  Authors of this model argue that the formation time of $D$ and
$B$ mesons, on the order of 1 fm, is much shorter then formation time of \piz\ particle and thus
the suppression pattern is due to the dissociation of fully formed charmed mesons in
contrast to \piz\ formed outside the interaction region.

Another surprise comes from the similarity of \raa\ measured for $\pi^{\pm}$ and
$p/\bar{p}$ particles \cite{Abelev:2006jr}. pQCD calculations indicate that a large fraction
of $p/\bar{p}$ baryons at high-\pt{}\ are produced from gluons unlike pions produced more by
valence quark interactions. Suppression factors for quark and gluon are expected to by
different because of different colored charge ($C_{A}/C_{F}$=9/4) \cite{Abelev:2007ra}.
However, the measured values of $\pi^{\pm}$ and $p\bar{p}$ \raa\ presented in
\cite{Abelev:2006jr} are almost identical in the \pt{}$\ge$5-6 \gevc.

The prompt photon measurement presented by the PHENIX experiment \cite{Isobe:2007ku} is
certainly one of the most interesting results at RHIC. At sufficiently high-\pt{}\ prompt
photons production is dominated by compton scattering ($q+g\rightarrow q+\gamma$) and to
some extent by $q\bar{q}$ annihilation, bremsstrahlung process and quark fragmentation.
Photons are blind to the final state interaction and thus no nuclear modification is
expected.  However, the preliminary data of PHENIX \cite{Isobe:2007ku} reveals quite
significant suppression in the \pt{}$\ge$15 \gevc\ region. Although the direct measurement
of nuclear shadowing at RHIC energies is not yet available, it is not likely that the
suppression of the prompt photon yield by a factor of two at \pt{}$\approx$18 \gevc\ could
be due to the shadowing \cite{Eskola:2001gt}. Also the isospin effect (colliding nuclei
contain protons and neutrons, however, nuclear data are compared to the yield in \pp\ and
it is known that photo-production cross section in \pp\ and $n+n$ are slightly different)
can explain some part the high-\pt{}\ suppression, but certainly not all.

It is evident that for deeper understanding of parton interaction with excited nuclear
medium an exploration of more exclusive processes like two particle, di-jet and
$\gamma_{direct}$-jet correlations is needed.

\section{Trigger-particle associated yields}

It has been discussed that one of the unsettled questions related to the high-\pt{}\
particle quenching is what fraction of energy partons are loosing via the radiative energy
loss and what fraction is purely collisional.  One of the possible ways to shed more light
on this question is to study the modification of the fragmentation function as suggested
\eg\ in \cite{Guo:2000nz,Borghini:2005em}.  The LHC era with yet larger hard-scattering
cross section opens the new regime of heavy ion physics allowing to explore the fully
reconstructed jets and their properties. However, due to the large particle multiplicity
produced in heavy ion collisions the reliable jet reconstruction is possible only at
relatively high transverse momenta above 50 \gev\ or more. In order to explore the
``low''-\pt{}\ region where the jet reconstruction is not possible the two-particle
correlation technique can be used.  However, extraction of underlaying parton properties
from two-particle correlation functions is not an easy task. For example, it has been
believed for decades that the shape of the high-\pt{}\ trigger hadron associated \xe\ distributions , where
\bge\label{eq:xe}
\xe=-{\vpt{t}\cdot\vpt{a}\over\pt{t}^2}=
\simeq-{\za\,\ptq{a}\over\zt\,\ptq{t}}
\ende
and \ptq{t}, \ptq{a}, \pt{t}, \pt{a} are the transverse momenta of the
trigger and assocated parton, trigger and associated hadrons and
\za=\pt{a}/\ptq{a} and \zt=\pt{t}/\ptq{t},  reflect the shape of the
fragmentation function. However, it has been demonstrated in \cite{longPRC:2006sc} that
the fixed momentum of the trigger particle \pt{t}\ does not fix the mother parton
momentum \ptq{t}\ and it varies with \pt{a}. In this case both \za\ and
\zt\ vary with \pt{a}\ and this variation completely masks the actual shape of the
fragmentation function.

One of the alternative ways to explore the fragmentation function is to study the particle
distributions associated to the prompt photon. In the high-\pt{}\ region where the
photo-production is dominated by the reverse Compton scattering diagram, the prompt photon
balances the back-to-back quark. In such a  case the measurement of the photon energy
determines also the outgoing quark energy and thus the measurement of transverse momentum
distribution of particles associated to the prompt photon corresponds  to the measurement
of the fragmentation function. However, also in this case there are important effects
which need to be taken into account. One constraint comes from the fact that at low \pt{}\
the prompt photon production may be contaminated by fragmentation photons from the large
number of gluonic jets produced in heavy ion collision \cite{Arleo:2007qw}. Fragmentation photons
obviously do not balance the away-side parton.

Another constraint comes from the fact that even when considering the leading order Compton
diagram there is always soft QCD radiation which breaks the jet-photon momentum balance and the
azimuthal collinearity. As a result of soft QCD radiation  This non-perturbative radiation manifests itself as a non-zero value of the net parton-photon pair transverse momentum magnitude \pt{pair}. 
Originally this soft radiation induced transverse momentum was attributed to the individual incoming partons
and the notation of \kt{}\ (parton transverse momentum) was introduced by Feynman, Field and Fox \cite{Feynman4}.

As discussed in \cite{longPRC:2006sc} the \kt{}\ may be viewed as a sum of three different contributions:
\bgi
\item Intrinsic parton Fermi-momentum \cite{Feynman4}.
\item Soft QCD radiation \cite{Werner_resummation}.
\item Hard NLO radiation.
\endi

\begin{figure}[h]
\begin{center}
  	\resizebox{8cm}{!}{\includegraphics{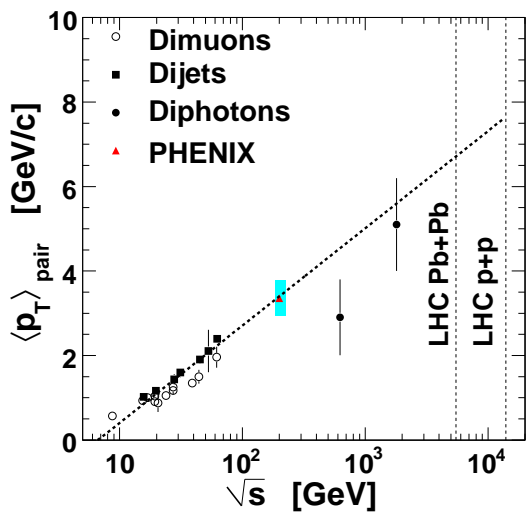}}
	\caption{
	The net dimuon (open circles), dijet (solid squares), diphoton (solid circles) and back-to-back partons 
	(red triangle) transverse momenta measured at various center of mass energies (see \cite
	{Apanasevich_kt_E609} and \cite{longPRC:2006sc}). PHENIX back-to-back parton value is derived from 
	dihadron accoplanarity when the average trigger hadron momentum fraction \zt\ is taken into account.
	 }
	\label{fig:ptpair}
\end{center}
\end{figure}
The magnitude of Fermi-momentum (truly intrinsic parton momentum) is determined by the
transverse size of colliding nucleons and it is of order of 300 \mevc. The large values of
\sqrtrms{\kt{}}$\approx$ 3 \gevc\ at RHIC indicates that the soft and hard radiation are
the dominant contributions. It evidently implies that it would be more appropriate to talk
about the net pair transverse momentum rather than about \kt{}\ attributed to individual
partons. However, we will stick to the historical notation of Feynman 
\footnote{in the case of 2D Gaussian distribution there is a trivial realtion \rms{\kt{}}=$\pi\over 2$\ptpair}
{\sl et al.}.

The net back-to-back parton transverse momentum \pt{pair}\ induced by three different
contributions discussed above will lead to the imbalance of the prompt photon and quark energies.
One way of overcoming the constraint is to trigger on photon momenta much larger than the
actual value of \kt{}. Extrapolation from measured values of \sqrtrms{\kt{}}\ to LHC
energy regime
$$
\ptpair\approx\log(0.15\cdot\sqrt{s})
$$
(see \fig{fig:ptpair}) leads to the values larger by factor of two (\ptpair=7.7 \gevc\ and
\sqrtrms{\kt{}}=6.1 \gevc\ at \s=14 \tev) as compared to RHIC. If this will be the case 
at LHC energy then for the unbiased region of prompt photon
momenta where the \ptpair\ will be negligible lies probably way above
\pt{t}$\ge$20\gevc.

\begin{figure}[h]
\begin{center}
  	\resizebox{8cm}{!}{\includegraphics[viewport=90 200 530 370,clip=]{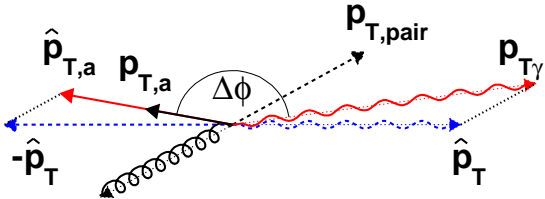}}
	\caption{ 
	Schematic view of Compton $q+g\rightarrow q+\gamma$ photo-production event in the
	plane perpendicular to the beam. In purely leading order and in the center of mass of
	the hard scattering, the outgoing photon and quark (blue arrows) are exactly
	back-to-back. The soft QCD radiation, represented by single gluon line, leads to
	the accoplanarity and imbalance between the photon, \ptg{}\, and quark, \ptq{a} momenta.
	 }
	\label{fig:jets}
\end{center} \end{figure}

In order to evaluate an effect of \kt{}\  imbalance more quantitatively let us consider the a cartoon of
hard scattering in the transverse plane as shown on \fig{fig:jets}. In the center of mass of 
the hard scattering and in purely leading order Compton $q+g\rightarrow q+\gamma$
the outgoing photon and quark are exactly back-to-back having momenta \ptq{} and -\ptq{}. 
However, due to the QCD radiation discussed earlier the quark+photon system acquire a boosted in transverse 
direction along the \pt{pair}\ vector and photon of momentum \ptg\ and quark \ptq{a}\ are emitted. 
Quark of momentum \ptq{a}\ fragments and particle of momentum \pt{a}\ is detected.
We assume that the invariant mass of the $q+\gamma$ system 
before and after the \pt{pair}\ boost is the same and thus 
$$
(\ptaf + \pttf)^2 = 2 \ptq{a}\ptg - 2  \ptq{a}\ptg \cos \Delta\phi = 4\ptq{}^2 \nonumber\\
$$

It is known that the magnitude of \pt{pair}\ has a Gaussian distribution. For example the 
muon pair transverse momentum distributions measured at $p_{\rm beam}$=400 \gevc\ fixed target experiment 
\cite{Ito:1980ev} have clearly Gaussian shape in \pt{pair}$\leq$3 \gevc\ with a hint of power law 
tail due to the NLO contribution at \pt{pair}$\geq$3 \gevc. Assuming the Gaussian distribution for 
\pt{pair}\ magnitude one can evaluate the production probability of quark-photon pair with \ptg\ and \ptq{a}\ 
given \ptpair\ and \ptq{} (see \fig{fig:jets}) as
\begin{eqnarray}\label{eq:probphot}
\prob(\ptq{a}\&\ptg)\Big|_{\ptq{}} =
\frac{\ptg + \ptq{a}}{\pi \sigma^2 \sqrt{\ptq{a}\ptg - \ptq{}^2}} \times\\
\times\exp\left[-\frac{(\ptg + \ptq{a})^2 - 4 \ptq{}^2}{2 \sigma^2}\right] \nonumber
\end{eqnarray}
where $\sigma^{2}$=\rms{\kt{}}. It is then straightforward to integrate \eq{eq:probphot} over \ptq{}\ 
(here we assumed as in \cite{longPRC:2006sc} dN/d\ptq{}$\propto\ptq{}^{-7}$) and 
fold with parametrized final state parton spectrum and fragmentation function as discussed in 
\cite{longPRC:2006sc} to evaluate the prompt photon associated spectra.

\begin{figure}[h]
\begin{center}
  	\resizebox{8cm}{!}{\includegraphics{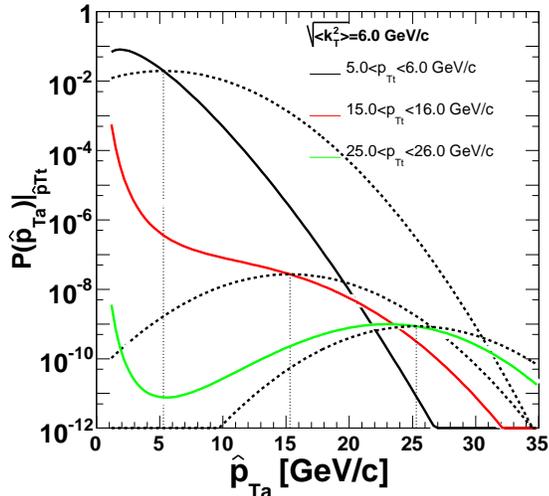}}
	\caption{ Associated parton distributions according integral of \eq{eq:probphot} to the 
	trigger photon of momentum \ptg=5, 10 and 25 \gevc\ (black, red and green solid lines)
	and \sqrtrms{\kt{}}=6 \gevc. Dashed lines represent the asymptotic Gaussian functions 
	discussed in the text. Vertical dashed lines represent the trigger photon energies.
	 }
	\label{fig:pta}
\end{center} \end{figure}

\fig{fig:pta} shows associated parton distributions calculated according \eq{eq:probphot} to the 
trigger photon of momentum \ptg=5, 10 and 25 \gevc\ and \sqrtrms{\kt{}}=6 \gevc. At high-\ptg\ 
region where $\ptg\gg\kt{}$ the associated parton distribution can be approximated by the Gaussian
function centered at \ptg\ and of variance $\sigma^{2}=\rms{\kt{}}/2$ (dashed curves on \fig{fig:pta}). 
More details can be found in \cite{longPRC:2006sc}.

\section{conclusion}
It is evident that at LHC one can expect that the imbalance due to the \kt{}\ smearing completely washes out the correlation between the quark and photon energies for the photon energy below 10 \gevc\ and the distributions deviate from the Gaussian
function quite significantly even at \ptg=15 \gevc. In order to extract the fragmentation function from the prompt photon 
associated distributions in the \ptg\ region below 30 \gevc\ at LHC the detailed understanding of  \kt{}\ phenomena is
necessary. 

\bibliographystyle{h-physrev3}
\bibliography{JanRak}

\end{document}